\documentclass{article}
\usepackage{arxiv}

\usepackage[utf8]{inputenc} 
\usepackage[T1]{fontenc}    
\usepackage[hidelinks]{hyperref}       
\usepackage{url}            
\usepackage{booktabs}       
\usepackage{amsfonts}       
\usepackage{nicefrac}       
\usepackage{microtype}      
\usepackage{graphicx}
\usepackage[capposition=below]{floatrow}

\usepackage{natbib}
\setcitestyle{authoryear,open={(},close={)}}

\usepackage{url}            

\usepackage{algorithm}
\usepackage{caption}
\usepackage{amsmath}
\usepackage{amsfonts}
\usepackage{amssymb}
\usepackage{calrsfs}
\usepackage{dsfont}

\newcommand{\mathdss}[1]{\text{\usefont{U}{dsrom}{m}{n}#1}}

\title{Topic Scaling: A Joint Document Scaling-Topic Model Approach To Learn Time-Specific Topics}

\author{
 Sami Diaf \\
  Faculty of Business, Economics and Social Sciences, Department of Socioeconomics\\
  Universität Hamburg\\
  \texttt{sami.diaf@uni-hamburg.de} \\
   \And
 Ulrich Fritsche \\
  Faculty of Business, Economics and Social Sciences, Department of Socioeconomics\\
  Universität Hamburg\\
  \texttt{ulrich.fritsche@uni-hamburg.de} \\
	}

\begin{document}

\maketitle  
\date{}            
\begin{abstract}
This paper proposes a new methodology to study sequential corpora by implementing a two-stage algorithm that learns time-based topics with respect to a scale of document positions and introduces the concept of \textit{Topic Scaling} which ranks learned topics within the same document scale. The first stage ranks documents using \textit{Wordfish}, a Poisson-based document scaling method, to estimate document positions that serve, in the second stage, as a dependent variable to learn relevant topics via a supervised Latent Dirichlet Allocation. This novelty brings two innovations in text mining as it explains document positions, whose scale is a latent variable, and ranks the inferred topics on the document scale to match their occurrences within the corpus and track their evolution. Tested on the U.S. State Of The Union two-party addresses, this inductive approach reveals that each party dominates one end of the learned scale with interchangeable transitions that follow the parties' term of office. Besides a demonstrated high accuracy in predicting in-sample documents' positions from topic scores, this method reveals further hidden topics that differentiate similar documents by increasing the number of learned topics to unfold potential nested hierarchical topic structures. Compared to other popular topic models, \textit{Topic Scaling} learns topics with respect to document similarities without specifying a time frequency to learn topic evolution, thus capturing broader topic patterns than dynamic topic models and yielding more interpretable outputs than a plain latent Dirichlet allocation.

\end{abstract}

\section{Introduction}

\subsection{Document scaling}

Document scaling refers to a popular class of methods, mostly unsupervised learning-based, used to study political manifestos and other corpora in social sciences as it creates a low-dimension metric to compare documents based on a set of assumptions regarding word distributions. Earlier approaches used machine learning algorithms, such as Na\"ive Bayes, to build scales \citep{McCallum1998}. \citet{wordscore} used \textit{Wordscores} to estimate political parties positions based on pre-established reference scores for texts. \citet{wordfish} designed \textit{Wordfish}, a parametric approach that uses a Poisson distribution model to infer a unidimensional document scale from the distribution of word frequencies, considered as a proxy for ideological positions. \textit{Wordshoal} \citep{wordshoal} is a two-stage document scaling method that applies \textit{Wordfish} on each debate of the corpus and then uses a Bayesian factor aggregation to uncover further dimensions related to the corpus metadata. 

As for \textit{Wordfish}, scaling techniques have been widely used in political science \citep{Grimmer2013}, particularly to study manifestos \citep{wordshoal} but as a dimensionality reduction technique, they come with the constraint of recovering just one of many possible hidden dimensions \citep{goet}, in addition to being unable to properly define its meaningfulness \citep{Grimmer2013}, suffering from possible word variations when handling corpora spanning over large periods \citep{goet} and sensitive to the chosen pre-processing steps when cleaning texts \citep{Denny2018}. \citet{goet} suggested the use of supervised approaches to have meaningful polarization results.

\subsection{Topic models}

Unsupervised topic models have been the usual choice for researchers working with text data, with the aim to unveil latent features or hidden structures of terms that explain the word-document occurrences. Latent Dirichlet Allocation (LDA) \citep{lda} is the traditional go-to algorithm for such exercises. This generative model assumes documents being mixtures of independent latent features called \textit{topics}, which are in their turn mixtures of words drawn from a Dirichlet distribution. Several other variants were later proposed to deal with specific cases \citep{Boyd-Graber2014} as to consider sequences, hierarchies and sentiments. Blei and Lafferty \citep{dtm} proposed a dynamic topic model (DTM) to study sequential corpora through a discrete-time variant of LDA whose architecture infers time-related topics, based on proportion priors drawn from a Markov process, suitable for big corpora spanning over a long time frame.

\citet{slda} proposed a supervised Latent Dirichlet Allocation (sLDA) that builds a generalized linear model (GLM) on top of a classic LDA to help inferring topics when documents are paired with labels. \citet{Boyd-Graber2010} built a multilingual model based on sLDA to capture how multilingual concepts are clustered into thematically coherent topics and how topics associated with text connect to an observed regression variable. Several other tools were designed as special cases of LDA, as for hierarchical topic models \citep{hlda}, Pachinko allocation models \citep{pachinko} and sentence labeling \citep{Lu2011} with the aim to uncover further latent structures in corpora based on hierarchies and specific structures. The number of topics to be learned, a hyperparameter, is set arbitrarily by users although a variety of methods were proposed to estimate it as for hierarchical Dirichlet process (HDP) \citep{hdp2011} which uses an online variational inference to uncover the number of topics in the collection of documents. \citet{Greene2017} used a two-layer non-negative matrix factorization model to explore topics dynamics in the corpus of European Parliament and found substantive niche topics not captured by standard dynamic topic models.  

Another class of topic models, Structural Topic Model, has been used to link exogenous variables with learned topics \citep{stm1} \citep{stm2} in order to investigate the impact of potential covariates on learned topics, by using document meta-data in the estimating step, to facilitate tracking the effect, variables could have on the learned topics.

Our approach builds a two-stage learning process: a document scaling (\textit{Wordfish}) on top of a supervised latent Dirichlet allocation, to allow uncovering time-dependent topics based on the position of each document in the corpus. Hence, even if the Dirichlet distribution is not sequential \citep{dtm}, topics could be learned with respect to the evolving distribution of word frequencies that served to estimate the latent scale which refers to document scores used as a time-based, dependent variable. We used \textit{Wordfish} for document scaling as it builds a unique scale for all documents, rather than Wordshoal which learns a distinct scale for each debate in the corpus \citep{wordshoal}.

To the best of our knowledge, there was no attempt in text mining to use scaling techniques beyond estimated document positions or explore further extensions using topic models. Our method is noticeably suited to study the dynamic structure of the corpus by uncovering potentially time-dependent, nested topics with the use of a measurement model (\textit{Wordfish}) for text data, rather than evolving topics that need further hyperparameter tuning as for setting the frequency (time stamp) for the analysis \citep{dtm}. The optimal number of topics in our method, still a hyperparameter, could be learned by maximizing a metric of choice, as for the root mean squared error (RMSE) between the estimated and predicted document positions or the log-likelihood of the estimated sLDA model. We notice that increasing the number of topics, despite slowing down the execution time for large corpora and not necessarily improving the RMSE accuracy when using regularization, helps uncovering the breadth of hidden topic structures similar to Pachinko Allocation Model \citep{pachinko} that are highly informative.

Our findings contribute to two distinctive fields of text mining. In document scaling, we try to explain document positions based on groups of words occurring together (topics) rather than individual words. In topic modeling, we learn these topics from a collection of documents that are, a priori, time-scaled regarding their word frequencies, so that two documents having similar scores will tend to have similar topic distributions. Lastly, the use of regularization allows increasing the number of topics without altering the predictive properties of the model and helps unfolding potential hierarchical structures in the learned topics.

In parallel of document scaling, this paper introduces the concept of \textit{topic scaling} which refers to a supervised method used to learn topics with respect to a labeled scale. This approach associates topic scores with document positions and ranks topics with their most occurrences in the corpus, to help tracking their distributions over time.

The remainder of the paper describes the two components upon which \textit{Topic Scaling} is built (section 2), then details the application results over the State Of The Union Corpus two-party corpus (section 3) and later compares the results with a plain LDA model, as well as a dynamic topic model. 

\subsection{State of The Union Addresses}

State Of The Union (SOTU) addresses bear an importance in politics \citet{Shogan2016} as the US Constitution (Article II, Section 3) requires the president to provide information to the Congress about \textit{measures as he shall judge necessary and expedient}, that is description of the current situation, priorities and the legislative intents \citep{Savoy2015a}. \citet{Petrocik2003} assumed that some party-related relationships between president could be seen in SOTU, as democratic tenures are more likely to deal with topics like education, family, welfare, and healthcare, while a republican presidency is frequently tied to free enterprise and business, reduction of expenses, or a support for the military sector. An alternative hypothesis would consider each president having his own priorities, independently from its predecessors, leading to a distinct vocabulary choice in his addresses \citet{Savoy2015a}. 

As a popular dataset in linguistics and text mining applications, SOTU has been studied by linguistics to investigate rhetoric or uncovering distinctive patterns, as for text clustering for presidential style \citep{Savoy2015a}, vocabulary-growth model \citep{Savoy2015b}, topic models of important words \citep{Savoy2016} or syntactic complexity \citep{Lei2020}. \citet{sotu4} studied the rhetorical changes of the SOTU addresses from George Washington until Bill Clinton and found three distinctive periods a founding, a traditional and a modern period, while \citet{Cummins2008} showed the importance of rhetorical attention to economic policy and foreign relations in modern addresses (1953-2000).

\section{Method}

This paper's algorithm, named \textit{Topic Scaling}, is a two-stage learning process for time-based topics from document positions, which serve as labels for supervised topic model. Hence, a sequential scale is built with \textit{Wordfish} model, and used as a label to run a supervised Latent Dirichlet Allocation that renders topic-allocation over time for historical corpora. This strategy combines two different machine learning approaches to give an augmented, automated content analysis method where documents and topics are put on a unique scale.

\begin{algorithm}
  \caption{Topic Scaling}
  \begin{enumerate}
    \item
    Estimate document positions $\hat{\psi}$ using \textit{Wordfish}

    \begin{enumerate}
      \item
			Assuming: $w_{ik} = \textrm{Poisson}(\lambda_{ik})$  

      \item
      Learn via Expected Maximization ($\alpha,\nu,\beta,\psi$) from: \\
			$log(\lambda_{ik}) =\alpha_{i}+\nu_{k}+\beta_{k} \times \psi_{i}$

    \end{enumerate}

  \item
  Learn a Supervised LDA with an L2 regularization (shrinkage parameter $\lambda$)
		\begin{enumerate}
      \item
			Draw topic proportions $\theta | \alpha \sim  Dir(\alpha)$ 
			\item
			For each word:
		\begin{enumerate}
      \item Draw a topic assignment: $z_{n}|\theta \sim Mult(\theta) $
      \item Draw word $w_{n}|z_{n},\beta_{1:K} \sim Mult(\beta_{z_{n}})  $
    \end{enumerate}
    \item 
		Draw document scale $\hat{\psi}|z_{1:N},\eta,\sigma^{2} \sim \mathcal{N}(\eta^\intercal \bar{z},\sigma^{2})$ with $z \sim \mathcal{N}(0,\frac{1}{\lambda})$
    \end{enumerate}
  \end{enumerate}
\end{algorithm} 

\subsection{Measurement model}

The measurement model of the \textit{Topic Scaling} consists of a parametric estimation of document positions that infers a scale to cluster documents based on word frequencies' similarities over time. \textit{Wordfish} \citep{wordfish} assumes word frequencies are drawn from a Poisson distribution, to estimate word effects and a hidden scale as proxy of document positions.
\begin{equation}
w_{ik} = \textrm{Poisson}(\lambda_{ik})  
\end{equation}
\begin{equation}
log(\lambda_{ik}) =\alpha_{i}+\nu_{k}+\beta_{k} \times \psi_{i}
\end{equation}

where: $\omega_{ik}$ is the count of the word $k$ in document $i$, $\lambda_{ik}$ is the parameter of the Poisson distribution denoting its mean and variance. Parameters to be learned are: $\alpha_{i}$ as a document-specific fixed effect, $\nu_{k}$ a word-specific fixed effect, $\beta_{k}$ is the relationship of word $k$ to the latent document position and $\psi_{i}$ is latent position of document $i$ or the measurement scale. Model is estimated via Expected Maximization algorithm, consisting of estimating word parameters ($\nu$, $\beta$) and document parameters ($\alpha$, $\psi$) alternatively, until reaching a convergence \citep{wordfish}. Variational inference (Monte Carlo-Markov Chains) with Bayesian priors could be used as well to estimate \textit{Wordfish} parameters. We notice that the main result of \textit{Wordfish}($\psi$) has an undetermined scale (direction) and needs to be tuned to identify the \textit{Wordfish} equation \citep{wordfish}. 

This parametric measurement allows a time series scale to be estimated solely based on the observed word frequencies. Although the model is prone to potential departures from the Poisson hypothesis of conditional independence (expectation being equal to the variance), it remains a robust method for outliers in word use \citep{wordshoal}, compared to Correspondence Analysis (CA) \citep{goet}.

\subsection{Supervised Topic model}

The Latent Dirichlet Allocation \citep{lda} could be seen as a Bayesian mixed-membership, unsupervised learning algorithm that learns independent structures, or groups of words, called topics from a collection of documents. Later, Blei and McAuliffe \citep{slda} proposed a supervised extension where the triplet documents-topics-words is kept under a generalized linear model (GLM) that accommodates a variety of response types, as mentioned in Figure 1, where for a dataset of observed document-response pairs $(w_{d,1:N},y_{d})$, the goal is to learn topic multinomials $\beta_{k}$, the Dirichlet parameter $\alpha$ of the topic proportions $\theta_{d}$ and the GLM parameters $\eta$ and $\sigma^{2}$, using likelihood estimation based on variational expectation-maximization \citep{slda}.

A supervised topic model \citep{slda} is used here as a second-stage method to learn topics from the corpus, where the dependent variable is the learned scale from the measurement model. An L2 regularization scheme on learned topic scores is used to allow overlapping topics and prevent hard clustering situations.

For the sake of illustration, we describe our approach as a two-stage model, but could also be interpreted as one-stage method as the Poisson model for document scaling is indeed a special case of the generalized linear model, used in the sLDA estimation \citep{slda}

\begin{figure}[h!]
\centering
\includegraphics[width=8cm]{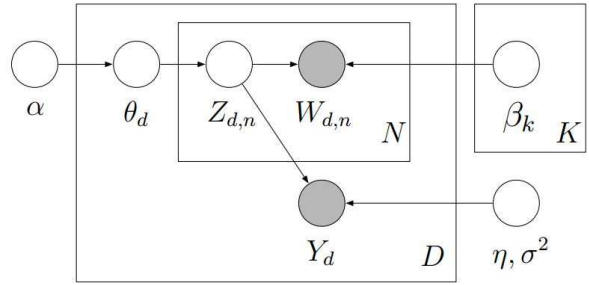}
\caption{Plate diagram of Supervised Latent Dirichlet allocation \citep{slda}.} \label{fig1}
\end{figure}

The number of topics to be learned could be approached via maximizing metrics related to sLDA as for $R^{2}$ or log-likelihood. \citet{Koltcov2020} proposed, based on statistical physics, a combination of renormalization procedure with the Rényi entropy for an efficient and quick estimation of the optimal number of topics in a given corpus. For $T$ topics and $N$ words in the corpus we set the deformation parameter $q=\frac{1}{T}$ and select $N$ words with a high probability ($\phi_{wt}>\frac{1}{W}$) in the topic-word matrix $\phi$, so to compute the density-of-states function $\rho=\frac{N}{WT}$. The energy can be expressed as:
 
\begin{equation}
E = -ln(\tilde{P}) = -ln \left( \frac{1}{T} \sum\limits_{w,t}^{} \phi_{wt} \mathdss{1}_{\phi_{wt}>\frac{1}{W}} \right) 
\end{equation}

with a partition function $Z_{q}=e^{-qE+S}=\rho\left(\tilde{P}\right)^{q}$. The Rényi entropy is defined as 

\begin{equation}
S_{q}^{R}=\dfrac{ln(Z_{q})}{q-1}=\dfrac{q \times ln(q\tilde{P})+ q^{-1} ln(\tilde{\rho})}{q-1}
\end{equation}

and its minimum corresponds to the optimal number of topics \citep{Koltcov2020}.

\section{Data and Results}

We use the corpus of the State Of The Union (SOTU) speeches available in the R package \textit{quanteda} \citep{data_sotu}, gathering 214 speeches of U.S. presidents from from 1790 until 2019. We keep, in our analysis, documents starting from 1853 to ensure a party duality democratic-republican in our corpus that later helps us studying document scaling variations, while words, previously lemmatized with \textit{spaCy} language model \citep{spacy}, having less than 3 occurrences in the selected corpus were excluded to reduce the size of the document term matrix for more efficiency.

\begin{figure}[h!]
\centering
\includegraphics[width=10cm,angle=270]{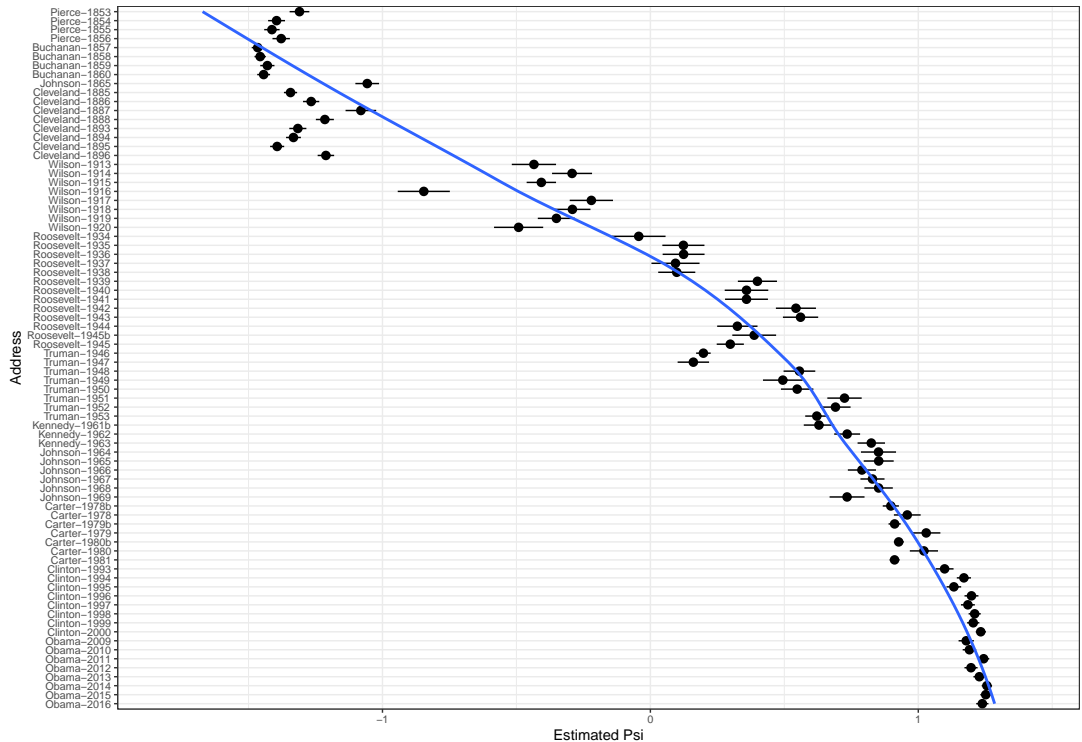}
\caption{\textit{Wordfish} scores for democratic presidents.} \label{fig2}
\end{figure}

\begin{figure}[h!]
\centering
\includegraphics[width=10cm,angle=270]{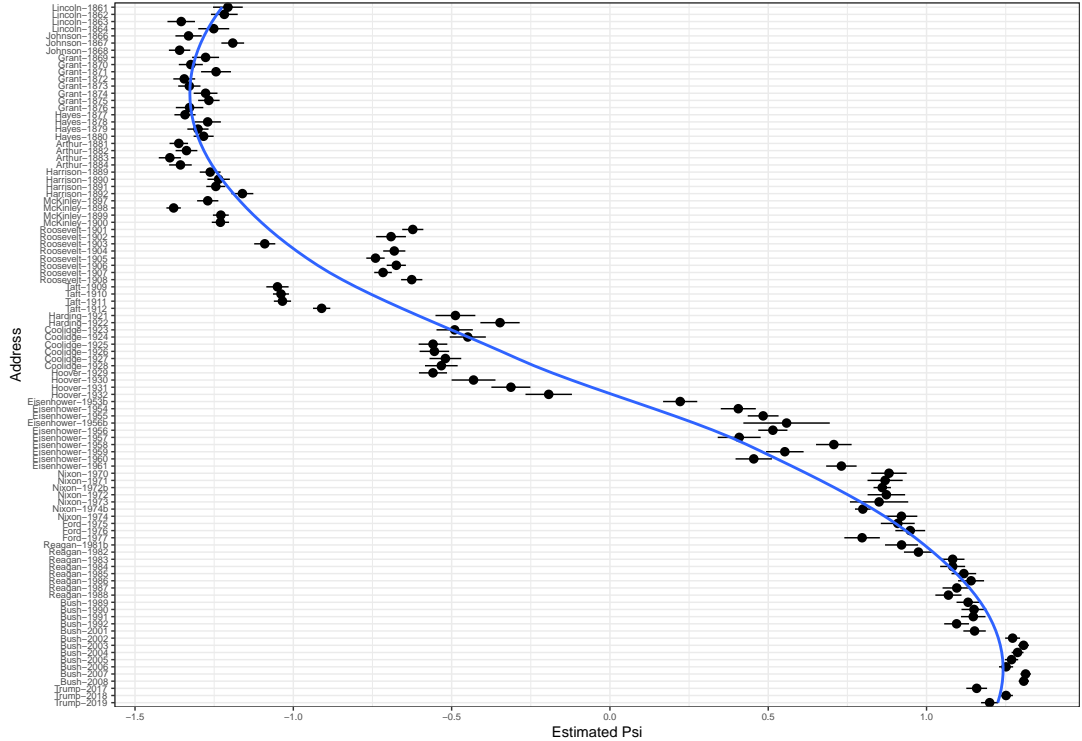}
\caption{\textit{Wordfish} scores for republican presidents.} \label{fig3}
\end{figure}

At first, document scaling (\textit{Wordfish}) was run on 176 documents, setting the score of Reagan's address in 1981 as being greater than Carter's one of the same year, to identify the scale parameter $\psi$. Figures 2 and 3 show \textit{Wordfish} results with a clear time-effect for both parties and noticeable similarities in addresses over time (in blue the \textit{Locally Weighted Scatterplot Smoothing} \textit{loess} curve \citep{loess1996}). Democratic addresses underwent a significant change in wording in earlier twentieth century with an exception of Wilson's speech of 1916 (Fig. 2), but modern addresses of Clinton and Obama share close document positions, and hence similarities, in line with the tree-based topical word-term representation of \citet{Savoy2015a}. At the other hand, republican addresses show greater variability in terms of document positions (Fig. 3) where it is common to have scattered positions of a single president, indicating a potential shift of interest in the addresses. 

Figure 4 shows the densities of document positions, by party, to be bimodal. Each party dominates one end of scale, with interchangeability linked to the presidents' tenure at office, as democratic addresses have a skewed distribution to the left, while republican ones have a less skewed one to the right. The document position $\hat{\psi}\approx$ 0.25, corresponding to the three addresses of president T.W. Wilson in 1914, could be interpreted as a cutoff or turning point that separates the studied corpus into two dual periods related to the evolution of rhetorical presidency \citep{sotu4}. \textit{Wordfish} scores confirm the results of \citet{Savoy2015a} who identified a distinctive style for each president since the speech of Roosevelt in 1934, while previous presidents shared many stylistic aspects.

\begin{figure}[h!]
\centering
\includegraphics[width=6cm,angle=270]{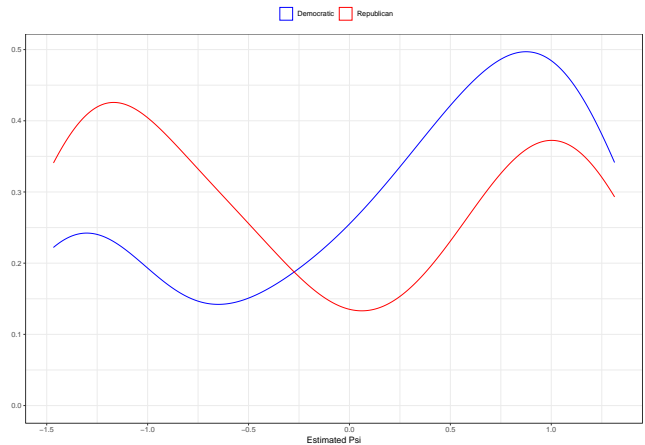}
\caption{Density plots of \textit{Wordfish} scores by party.} \label{fig4}
\end{figure}

Word positions plot (Fig. 5) shows the contribution of specific words to the estimated \textit{Wordfish} scale (words with higher $|\beta_{k}|$) and their specific effects ($\nu_{k}$). It appears that words related to security, mostly found in modern speeches, contribute mostly to documents with positive scores, that is recent addresses, even if their specific effects are relatively low. 

\begin{figure}[h!]
\centering
\includegraphics[width=8cm,angle=270]{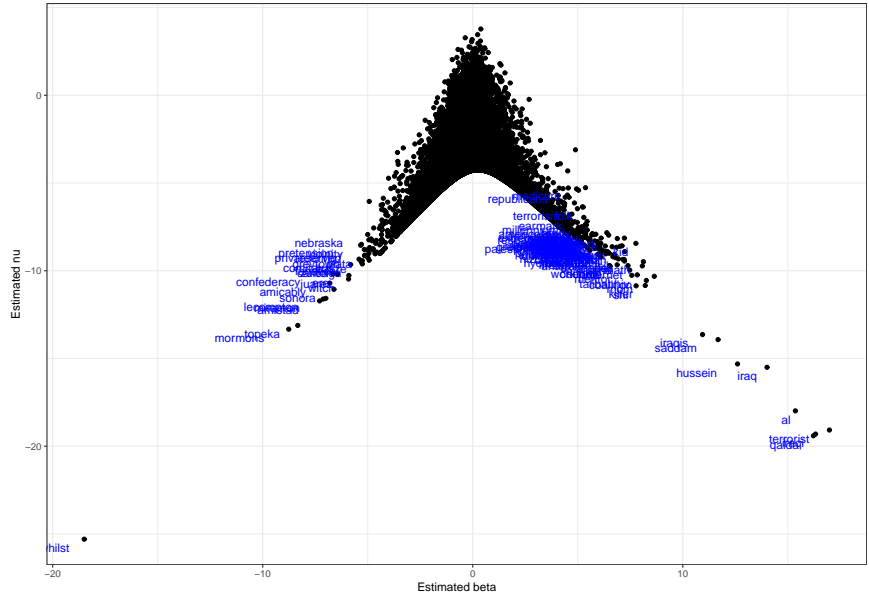}
\caption{Estimated word positions from \textit{Wordfish}.} \label{fig5}
\end{figure}

In the second stage, supervised LDA was run through Expected Maximization algorithm (50 steps expectation and 20 steps maximization), with $\alpha=1$ (Dirichlet hyperparameter for topic proportions), $\eta=0.1$ (Dirichlet hyperparameter for topic multinomials) and an L2 regularization scheme to learn topic scores with a shrinkage parameter $\lambda=0.01$. This setting was tested against a plain LDA with as well as a dynamic topic model by keeping the same hyperparameter setting used for sLDA ($\alpha=1$ and $\eta=0.1$).

\begin{figure}[h!]
\centering
\includegraphics[width=9cm,angle=270]{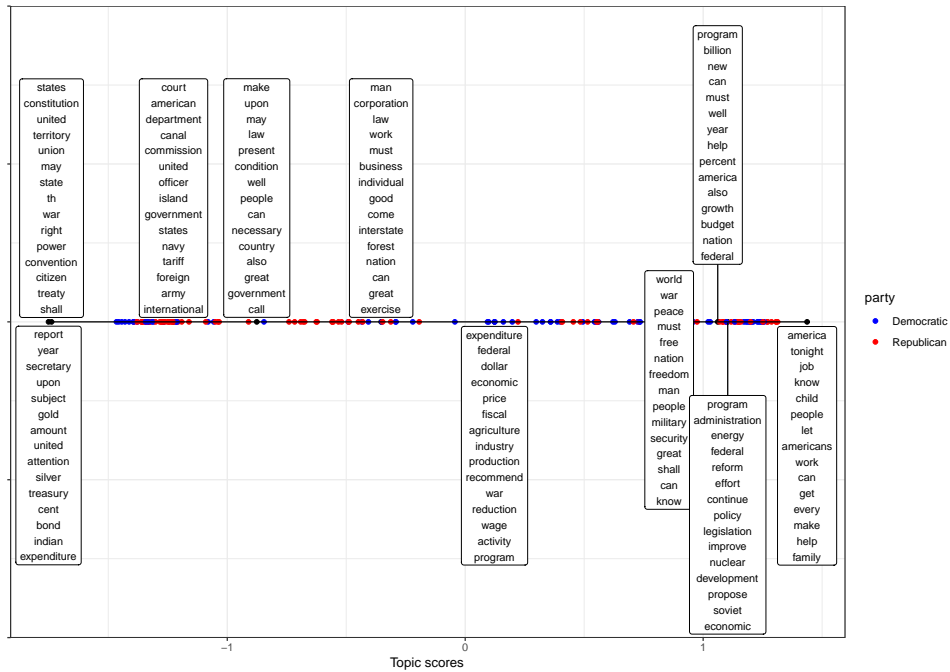}
\caption{Topic scores learned via \textit{Topic Scaling} (10 topics).} \label{fig6}
\end{figure}

\begin{figure}[h!]
\centering
\includegraphics[width=8.5cm,angle=270]{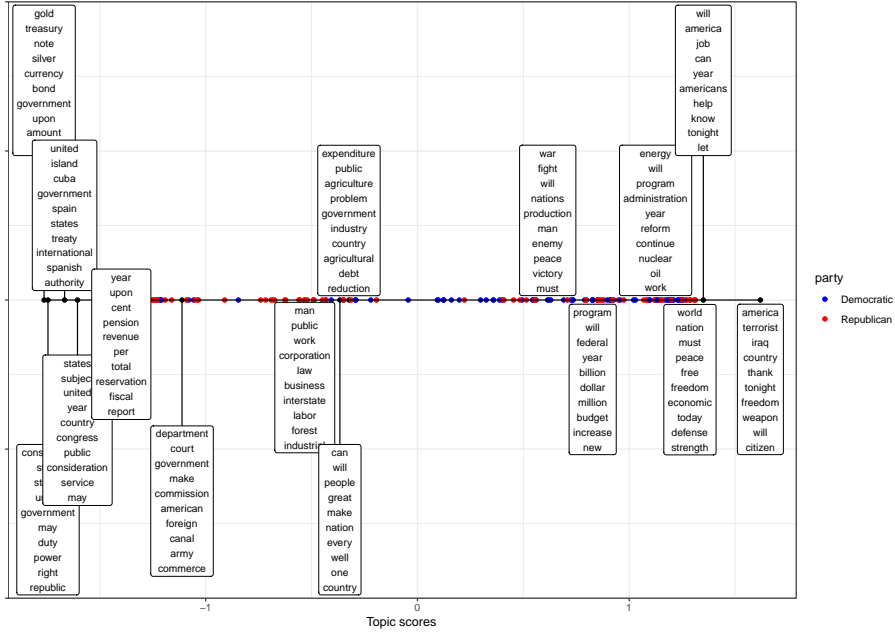}
\caption{Topic scores learned via \textit{Topic Scaling} (15 topics).} \label{fig7}
\end{figure}

Figures 6 and 7 show top 10 words in each learned \textit{Topic Scaling} model with 10 and 15 topics respectively and confirms the existence of nested structures when increasing the number of topics. Model with 15 topics seems to be the winning solution as it maximizes the Rényi entropy (Table 1), while other metrics ($R^{2}$ and log-likelihood) were not found to be informative with increasing number of topics.

Documents with negative scores are likely to belong to the second half of the ninetieth century and the first three decades of the twentieth century, where topics are related to both domestic and international environments with an emphasis on government affairs (regulation and administration), while topics with positive scores are linked to modern addresses, focusing on economic and security issues. These two distinct windows differentiate topics and are separated by the period where addresses occurred in 1930s, in-line with Fig. 4 which indicates two intertwined regimes in document scores.

Recent addresses favor economic welfare, government affairs and security and international environment, corresponding to the 4 topics in the left of the scale (Fig. 6) which seem to be party-related topics \citep{Savoy2015a}. Increasing the number of topics to 15 (Fig. 7) reveals further sub-topics that provide a better area-specific understanding as for terrorism, foreign policy, internal affairs and labor. A few number of topics will cluster the topic content so that topic interpretation cannot be easily given, while increasing the topics unveil clearer policy fields of the addresses that help distinguishing similarities in the addresses within a moderate time frame. 
     
As a matter of comparison, results of a plain LDA with 15 topics (Table 3) and a dynamic topic model with 3 per-decade topics (Table 4) do not yield relevant structures, compared to our method. A clear dominance of frequent words used in the rhetoric (as for \textit{government}, \textit{will}, \textit{administration}, \textit{Congress}) is seen in almost all topics, reducing the model's informative content.

 The \textit{scale-effect}, referring to the use of \textit{Wordfish} scores, helps distinguishing topics by periods where documents present certain similarities in the distribution of word frequencies. One could interpret the learned topics as being the most dominant probability distribution of words over a specific time frame, where addresses exhibit similar word usage.

Finally, \textit{Topic Scaling} displays interesting findings in terms of topics' contribution. Two topics (Topics 5 and 6) are forming the building blocks of modern addresses and are seen as dominant topics in recent speeches (Fig. 8 and 9), dealing family, economic condition and foreign affairs.

\begin{figure}
\centering
\includegraphics[width=6cm,angle=270]{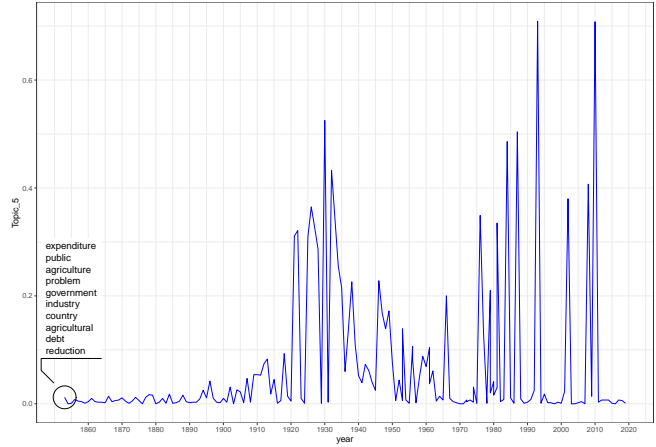}
\caption{Evolution of \textit{Topic 5} proportions in SOTU speeches (\textit{Topic Scaling}).} \label{fig8}
\end{figure}

\begin{figure}
\centering
\includegraphics[width=6cm,angle=270]{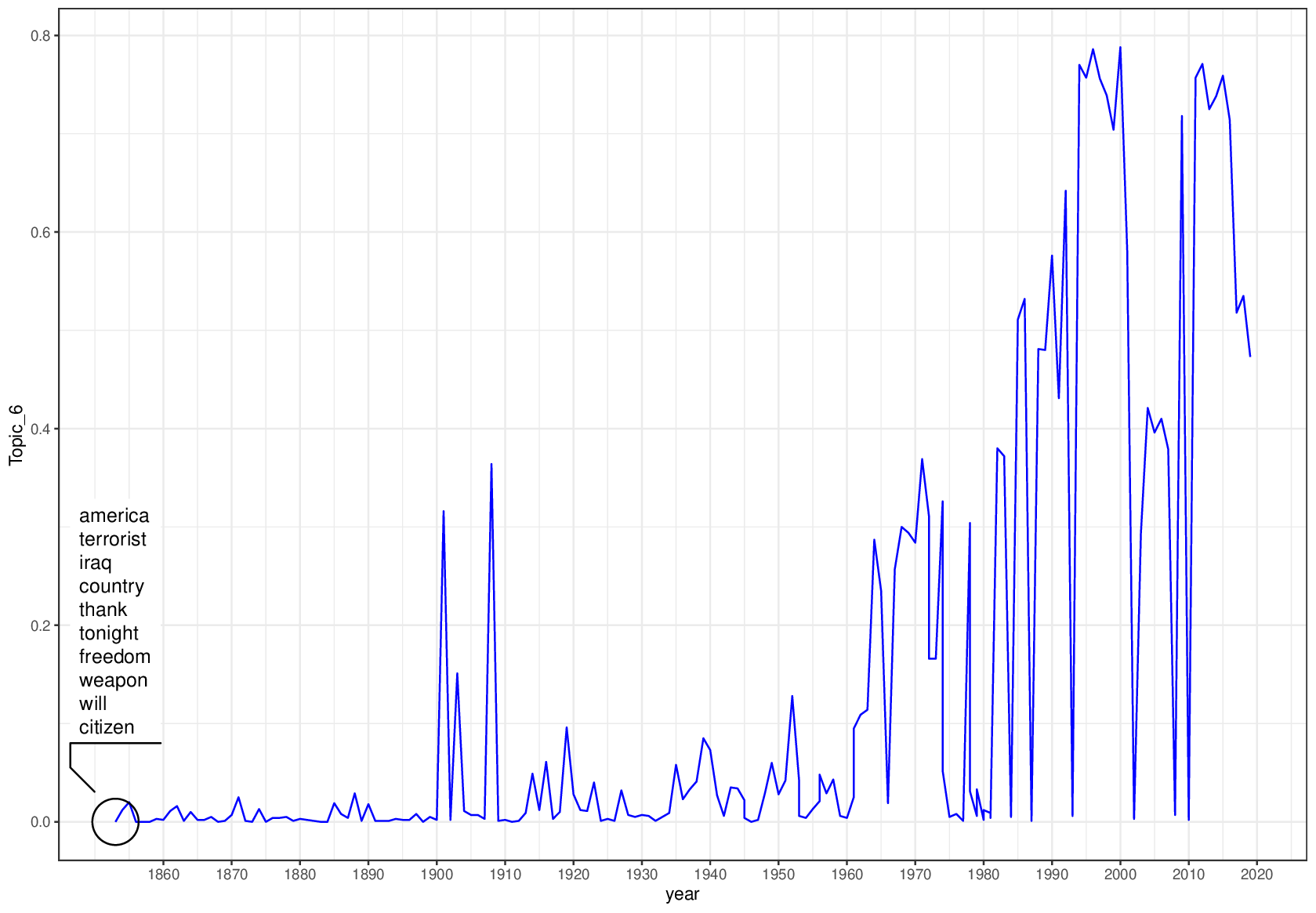}
\caption{Evolution of \textit{Topic 6} proportions in SOTU speeches (\textit{Topic Scaling}).} \label{fig9}
\end{figure}

\begin{table}[h!]
\centering
\scalebox{0.8}{
\begin{tabular}{|c|c|c|c||c|c|c|c|}
  \hline
Number of Topics & $R^{2}$ & Log-Likelihood  & Rényi entropy & Number of Topics & $R^{2}$ & Log-Likelihood  & Rényi entropy\\ 
  \hline
\hline
4 & 0.9963 & 236.5854 & -3.7986 & 15 & 0.9976 & 287.9229 & -3.3353 \\
\hline
5 & 0.9946 & 207.1414 & -3.6783 & 16 & 0.9975 & 276.7568 & -3.4440 \\
\hline
6 & 0.9964 & 246.9864 & -3.6557 & 17 & 0.9969 & 261.3767 & -3.3559 \\
\hline
7 & 0.9962 & 240.2083 & -3.6772 & 18 & 0.9978 & 281.0514 & -3.3979 \\
\hline
8 & 0.9957 & 228.3746 & -3.5676 & 19 & 0.9980 & 293.1175 & -3.4628 \\
\hline
9 & 0.9963 & 235.5375 & -3.5626 & 20 & 0.9981 & 298.9496 & -3.5025 \\
\hline
10 & 0.9963 & 245.5822 & -3.3909 & 21 & 0.9978 & 295.9086 & -3.5367 \\ 
\hline
11 & 0.9977 & 273.5398 & -3.4583 & 22 & 0.9976 & 272.1873 & -3.5247 \\
\hline
12 & 0.9968 & 255.8592 & -3.4838 & 23 & 0.9980 & 285.0266 & -3.6266 \\
\hline
13 & 0.9976 & 279.4950 & -3.3901 & 24 & 0.9981 & 301.2945 & -3.5919 \\
\hline
14 & 0.9967 & 245.8567 & -3.4175 & 25 & 0.9977 & 293.4041 & -3.6431 \\
\hline
\end{tabular}}
\caption{Results of \textit{Topic Scaling} metrics, per different number of topics.}
\end{table}

\begin{table}[h!]
\centering
\scalebox{0.9}{
\begin{tabular}{|c|l|}
\hline
Topic & Top 10 Words\\
\hline
1 & energy,will,program,administration,year,reform,continue,nuclear,oil,work\\
\hline
2 & united,island,cuba,government,spain,states,treaty,international,spanish,authority\\
\hline
3 & program,will,federal,year,billion,dollar,million,budget,increase,new\\
\hline
4 & america,terrorist,iraq,country,thank,tonight,freedom,weapon,will,citizen\\
\hline
5 & expenditure,public,agriculture,problem,government,industry,country,agricultural,debt,reduction\\
\hline
6 & will,america,job,can,year,americans,help,know,tonight,let\\
\hline
7 & man,public,work,corporation,law,business,interstate,labor,forest,industrial\\
\hline
8 & department,court,government,make,commission,american,foreign,canal,army,commerce\\
\hline
9 & war,fight,will,nations,production,man,enemy,peace,victory,must\\
\hline
10 & states,subject,united,year,country,congress,public,consideration,service,may\\
\hline
11 & constitution,state,states,union,government,may,duty,power,right,republic\\
\hline
12 & can,will,people,great,make,nation,every,well,one,country\\
\hline
13 & world,nation,must,peace,free,freedom,economic,today,defense,strength\\
\hline
14 & year,upon,cent,pension,revenue,per,total,reservation,fiscal,report\\
\hline
15 & gold,treasury,note,silver,currency,bond,government,upon,amount,duty\\
\hline
\end{tabular}}
\caption{Top 10 words from the estimated \textit{Topic Scaling} with 15 topics.}
\end{table}

\begin{table}[h!]
\centering
\scalebox{0.9}{
\begin{tabular}{|c|l|}
\hline
Topic & Top 10 Words\\
\hline
1 & will, america, must, world, nation, can, year, people, help, freedom\\
\hline
2 & will, war, world, nation, can, must, people, great, man, peace\\
\hline
3 & will, program, year, must, government, can, nation, congress, federal, new\\
\hline
4 & government, will, year, make, war, congress, country, can, federal, public\\
\hline
5 & man, law, will, government, make, can, great, nation, people, work\\
\hline
6 & will, year, program, congress, federal, administration, new, increase, continue, energy\\
\hline
7 & states, united, government, congress, year, may, will, country, upon, make\\
\hline
8 & will, upon, make, year, law, people, government, country, public, great\\
\hline
9 & government, states, united, year, will, make, upon, congress, american, may\\
\hline
10 & states, government, united, state, may, congress, will, power, constitution, upon\\
\hline
11 & will, year, must, work, people, can, child, america, new, make\\
\hline
12 & will, year, can, america, people, new, american, great, congress, nation\\
\hline
13 & government, make, will, states, united, congress, department, american, year, law\\
\hline
14 & government, upon, condition, may, present, year, law, make, gold, time\\
\hline
15 & will, year, can, job, make, work, america, people, new, american\\
\hline
\end{tabular}}
\caption{Top 10 words from an estimated plain LDA with 15 topics.}
\end{table}

\begin{flushleft}
\begin{table}[h!]
\centering
\scalebox{0.5}{
\begin{tabular}{|l|l|l|l|l|l|l|l|l|l|l|l|l|l|l|l|}

\hline
Decade & \multicolumn{3}{c|}{1853-1859} & \multicolumn{3}{c|}{1860-1869} & \multicolumn{3}{c|}{1870-1879} & \multicolumn{3}{c|}{1880-1889} & \multicolumn{3}{c|}{1890-1899}\\
\hline
Topics & 1 & 2 & 3 & 1 & 2 & 3 & 1 & 2 & 3 & 1 & 2 & 3 & 1 & 2 & 3\\
\hline
 & states & constitution & people & states & constitution & people & states & government & people & states & government & people & government & government & people\\

 & government & would & great & government & would & great & government & would & great & government & would & great & states & would & great\\

 & united & government & nation & united & government & nation & united & people & nation & united & people & nation & united & people & nation\\

 & congress & states & would & congress & states & would & congress & constitution & peace & congress & states & world & congress & public & world\\

 & would & people & peace & would & people & peace & would & states & world & would & public & peace & would & states & peace\\

 & country & country & world & country & country & world & country & country & would & country & country & every & great & country & every\\

 & public & congress & every & public & congress & every & public & public & every & public & congress & would & country & congress & would\\

 & great & public & power & great & public & power & great & congress & government & great & constitution & government & general & present & government\\

 & people & power & government & people & power & government & people & power & nations & people & present & nations & service & national & nations\\

 & citizens & state & nations & citizens & state & nations & general & present & power & general & power & power & people & business & national\\
\hline
\end{tabular}}
\end{table}

\begin{table}[h!]
\centering
\scalebox{0.5}{
\begin{tabular}{|l|l|l|l|l|l|l|l|l|l|l|l|l|l|l|l|}
\hline
Decade & \multicolumn{3}{c|}{1900-1909} & \multicolumn{3}{c|}{1910-1919} & \multicolumn{3}{c|}{1920-1929} & \multicolumn{3}{c|}{1930-1939} & \multicolumn{3}{c|}{1940-1949}\\
\hline
Topics & 1 & 2 & 3 & 1 & 2 & 3 & 1 & 2 & 3 & 1 & 2 & 3 & 1 & 2 & 3\\
\hline
 & government & government & people & government & government & people & government & government & world & government & government & world & government & government & world\\

 & states & would & great & states & would & world & states & congress & people & states & congress & people & congress & federal & people\\

 & united & public & nation & united & public & nation & congress & would & nation & congress & federal & nation & states & congress & nation\\

 & congress & congress & world & congress & congress & great & united & public & great & united & public & peace & united & program & peace\\

 & would & people & peace & would & country & peace & would & federal & peace & would & would & nations & would & public & nations\\

 & great & country & every & american & states & every & american & national & nations & country & national & great & country & legislation & government\\

 & american & states & would & great & people & government & great & country & government & american & legislation & government & great & national & congress\\

 & service & national & government & country & national & would & country & states & would & great & states & congress & service & would & great\\

 & country & business & nations & service & business & nations & service & people & every & service & country & national & american & states & national\\

 & general & present & national & department & federal & national & department & legislation & national & department & program & would & department & administration & economic\\
\hline
\end{tabular}}
\end{table}
\end{flushleft}

\vspace{-2em}

\begin{table}[h!]
\centering
\scalebox{0.5}{
\begin{tabular}{|l|l|l|l|l|l|l|l|l|l|l|l|l|}
\hline
Decade & \multicolumn{3}{c|}{1950-1959} & \multicolumn{3}{c|}{1960-1969} & \multicolumn{3}{c|}{1970-1979} & \multicolumn{3}{c|}{1980-1989} \\
\hline
Topics & 1 & 2 & 3 & 1 & 2 & 3 & 1 & 2 & 3 & 1 & 2 & 3 \\
\hline
 & government & federal & world & government & federal & world & congress & federal & world & congress & administration & world\\

 & congress & government & people & congress & program & people & government & administration & people & government & federal & people\\

 & states & congress & nation & states & government & congress & states & program & congress & states & program & congress\\

 & united & program & nations & united & congress & nation & united & congress & years & energy & congress & years\\

 & would & legislation & peace & would & administration & years & energy & legislation & nation & united & legislation & america\\

 & country & administration & congress & energy & legislation & government & administration & government & government & administration & development & american\\

 & service & public & government & administration & development & peace & would & development & america & would & government & government\\

 & great & national & years & country & national & nations & legislation & policy & american & legislation & policy & nation\\

 & american & development & economic & legislation & policy & economic & foreign & national & peace & foreign & national & programs\\

 & department & would & great & service & public & american & country & states & economic & country & assistance & economic\\
\hline
\end{tabular}}
\end{table}

\vspace{-2em}

\begin{table}[h!]
\centering
\scalebox{0.5}{
\begin{tabular}{|l|l|l|l|l|l|l|l|l|l|l|l|l|}
\hline
Decade & \multicolumn{3}{c|}{1990-1999} & \multicolumn{3}{c|}{2000-2009} & \multicolumn{3}{c|}{2010-2019}  \\
\hline
Topics & 1 & 2 & 3 & 1 & 2 & 3 & 1 & 2 & 3  \\
\hline
 & congress & administration & people & congress & administration & america & congress & administration & america\\

 & government & federal & america & government & federal & people & government & federal & people\\

 & states & program & world & states & program & american & states & program & american\\

 & energy & congress & years & energy & congress & years & energy & legislation & years\\

 & united & legislation & congress & united & legislation & world & united & congress & world\\

 & administration & development & american & administration & development & congress & administration & development & americans\\

 & would & government & government & would & government & americans & would & government & congress\\

 & legislation & policy & nation & legislation & policy & every & american & policy & every\\

 & american & national & americans & american & national & government & legislation & national & country\\

 & country & assistance & every & country & states & nation & country & states & tonight\\
\hline
\end{tabular}}
\caption{Top 10 words from an estimated Dynamic Topic Model (per decade) with 3 topics.}
\end{table}

\section{Conclusion}

We present a novelty in text mining, suited to study sequential corpora and outperforming other topic models in terms of interpretation and parametrization. \textit{Topic Scaling} could be seen as a dual algorithm: a supervised scaling method where topics are scaled on the same ideological dimension of documents, and a robust alternative to other sequential topic models where the estimated document scores serve as an ordered variable to retrieve topics rather than a learning process requiring a time frame. Under regularization schemes and entropy-related metrics, increasing the number of topics helps maximizing the information gain and uncovering nested structures that render information about potential embedded subtopics, thus, unveiling topics that signal important changes in the evolution of the corpus. Applied to study the party duality democratic-republican in the State Of The Union addresses, this method confirms the existence of two distinct periods correlated with the prevailing topics throughout the modern history of the Unites States, with a clear dominance of foreign affairs and business discourse in post-war addresses, while recent addresses seem to prioritize security and the economic status. Results confirm previous findings about the rhetoric as for vocabulary-rich, varying modern speeches and party-related subject preferences expressed in earlier addresses.

\bibliographystyle{unsrt} 
\bibliography{my_biblio}

\end{document}